# The Fluorescence detector Array of Single-pixel Telescopes: Contributions to the 35th International Cosmic Ray Conference (ICRC 2017)

**Contributions**

1. **"First results from the full-scale prototype for the Fluorescence detector Array of Single-pixel Telescopes"**, Toshihiro Fujii for the FAST Collaboration

2. **"The Prototype Opto-mechanical System for the Fluorescence detector Array of Single-pixel Telescopes"**, Dusan Mandat for the FAST Collaboration

# First results from the full-scale prototype for the Fluorescence detector Array of Single-pixel Telescopes

**Toshihiro Fujii**[*,a], **Max Malacari**[b], **Justin Albury**[c], **Jose A. Bellido**[c], **John Farmer**[b], **Aygul Galimova**[b], **Pavel Horvath**[d], **Miroslav Hrabovsky**[d], **Dusan Mandat**[e], **Ariel Matalon**[b], **John N. Matthews**[f], **Maria Merolle**[b], **Xiaochen Ni**[b], **Libor Nozka**[d], **Miroslav Palatka**[e], **Miroslav Pech**[e], **Paolo Privitera**[b], **Petr Schovanek**[e], **Stan B. Thomas**[f], **Petr Travnicek**[e] **(FAST Collaboration)**[†]

[a] *Institute for Cosmic Ray Research, University of Tokyo, Kashiwa, Chiba, Japan*
[b] *Kavli Institute for Cosmological Physics, University of Chicago, Chicago, IL, USA*
[c] *Department of Physics, University of Adelaide, Adelaide, S.A., Australia*
[d] *Palacky University, RCPTM, Olomouc, Czech Republic*
[e] *Institute of Physics of the Academy of Sciences of the Czech Republic, Prague, Czech Republic*
[f] *High Energy Astrophysics Institute and Department of Physics and Astronomy, University of Utah, Salt Lake City, UT, USA*
E-mail: fujii@icrr.u-tokyo.ac.jp

The Fluorescence detector Array of Single-pixel Telescopes (FAST) is a design concept for the next generation of ultrahigh-energy cosmic ray (UHECR) observatories, addressing the requirements for a large-area, low-cost detector suitable for measuring the properties of the highest energy cosmic rays. In the FAST design, a large field of view is covered by a few pixels at the focal plane of a mirror or Fresnel lens. Motivated by the successful detection of UHECRs using a prototype comprised of a single 200 mm photomultiplier-tube and a 1 $m^2$ Fresnel lens system [Astropart.Phys. 74 (2016) 64-72], we have developed a new full-scale prototype consisting of four 200 mm photomultiplier-tubes at the focus of a segmented mirror of 1.6 m in diameter. In October 2016 we installed the full-scale prototype at the Telescope Array site in central Utah, USA, and began steady data taking. We report on first results of the full-scale FAST prototype, including measurements of artificial light sources, distant ultraviolet lasers, and UHECRs.

---

[*]Speaker.
[†]https://www.fast-project.org

## 1. Fluorescence detector Array of Single-pixel Telescopes (FAST)

The origin and nature of ultrahigh-energy cosmic rays (UHECRs) is one of the most intriguing mysteries in particle astrophysics [1]. Given their minute flux, less than one particle per century per square kilometre at the highest energies, a very large area must be instrumented to collect significant statistics. The energy spectrum, arrival directions, and mass composition of UHECRs can be inferred from studies of the cascade of secondary particles (extensive air shower, EAS) produced by their interaction with the Earth's atmosphere. Two well-established techniques are used for UHECR detection: 1) arrays of detectors (e.g. plastic scintillators and water-Cherenkov stations) sample EAS particles reaching the ground; 2) large field of view telescopes allow for reconstruction of the shower development in the atmosphere by imaging ultraviolet (UV) fluorescence light from atmospheric nitrogen excited by EAS particles.

The Pierre Auger Observatory (Auger) [2] and the Telescope Array Experiment (TA) [3, 4], the two largest UHECR experiments currently in operation, combine the two techniques, with arrays of particle detectors overlooked by fluorescence detector (FD) telescopes. Auger covers an area of over 3000 km$^2$ close to the town of Malargüe in the province of Mendoza, Argentina. TA is located near the town of Delta in central Utah, USA, and covers an area of 700 km$^2$. Significant advances in our understanding of UHECRs have been achieved in the last decade by these experiments [5]. However, these results are limited by low statistics at the highest energies. To further advance the field, the next generation of experiments will require an aperture which is larger by an order of magnitude. This may be accomplished by fluorescence detection of UHECR showers from space, as in the proposed JEM-EUSO mission [6], or with a giant ground array. Low-cost, easily-deployable detectors will be essential for a future ground-based experiment.

We present a ground-based FD telescope concept which would fulfill these requirements. The Fluorescence detector Array of Single-pixel Telescopes (FAST) would consist of compact FD telescopes featuring a smaller light-collecting area and far fewer pixels than current-generation FD designs, leading to a significant reduction in cost. In the FAST design, a 30° × 30° field of view is covered by just a few 200 mm photomultiplier-tubes (PMTs) at the focal plane of a mirror or Fresnel lens with a ∼ 1 m$^2$ aperture. FAST stations, powered by solar panels and with wireless connection, could be deployed in an array configuration to cover a very large area at low cost.

## 2. The full-scale FAST prototype

A first test of the FAST concept was performed in 2014 using a single 200 mm PMT at the focus of a 1 m$^2$ Fresnel lens system at the Telescope Array site. Using the first prototype we detected 16 highly significant UHECR shower signals, and demonstrated excellent operational stability under conditions typical of field deployment (changes in temperature, night sky background and atmosphere; airplanes in the field of view; unexpected power cuts) [7]. Motivated by these encouraging results, we have developed a full-scale FAST prototype. The new prototype, shown in Figure 1, consists of a segmented spherical mirror of 1.6 m diameter (produced at the Joint Laboratory of Optics in Olomouc, Czech Republic), and a UV band-pass filter (ZWB3, Shijiazhuang Zeyuan Optics) with a 1 m$^2$ aperture. Four 200 mm PMTs (mod. R5912-03, Hamamatsu) and active bases (mod. E7694-01, Hamamatsu) are installed at the focal plane of the segmented mirror in a 2 × 2 matrix, covering a 25° × 25° field of view. The telescope frame is covered with a shroud to shield the optical system from dust and stray light. Details of the optical design of this prototype are presented at this conference [8].

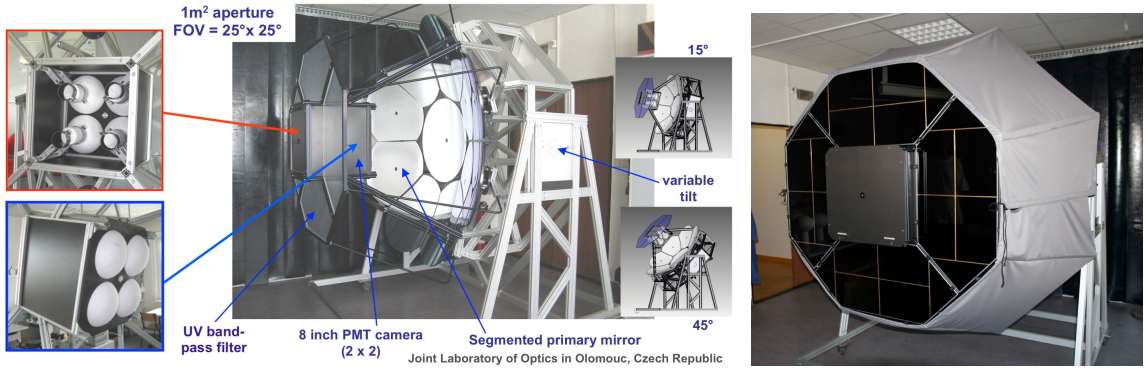

**Figure 1:** The full-scale FAST prototype developed at the Joint Laboratory of Optics in Olomouc, Czech Republic.

## 3. Installation of the full-scale FAST prototype and the first light operation

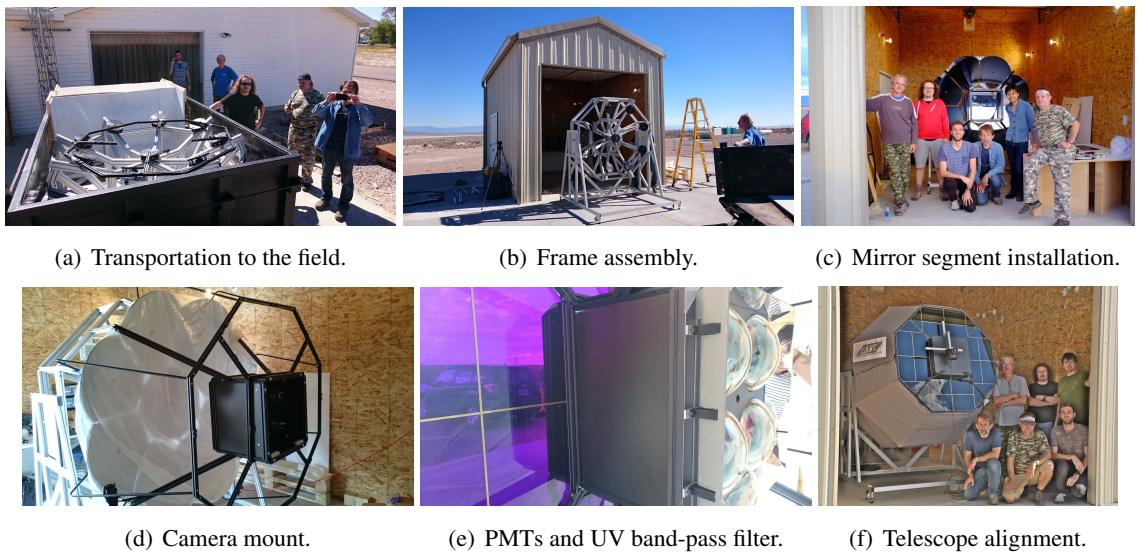

(a) Transportation to the field. (b) Frame assembly. (c) Mirror segment installation.

(d) Camera mount. (e) PMTs and UV band-pass filter. (f) Telescope alignment.

**Figure 2:** Photographs taken during the installation of the full-scale FAST prototype at the Telescope Array site.

In October 2016 the full-scale FAST prototype was installed at the Telescope Array site. Some photographs taken during the installation of the full-scale FAST prototype are shown in Figure 2. The telescope frame was assembled on site, before the PMTs were mounted in the camera box and the UV band-pass filter was installed at the telescope aperture. The telescope was aligned astrometrically using a camera mounted to the exterior of the frame [8]. Following its installation, first light operation of the prototype began, utilizing external air shower triggers from the adjacent TA fluorescence detector.

High voltage is supplied to the four PMTs, which were calibrated in the laboratory [9] to a nominal gain of $5 \times 10^4$, by a high-voltage power supply (mod. N1470, CAEN). The PMT signals are amplified by a factor of 50 using a fast amplifier (mod. 777, Phillips Scientific), and passed through a 15 MHz low-pass filter before digitization by a 12-bit FADC (mod. SIS3350, Struck Innovative Systeme) at a 50 MHz sampling rate. The digitizer is hosted in a portable VME crate (mod. VME8004B, CAEN), together with a controller (mod. V7865, GE Intelligent platforms) and a GPS unit (mod. GPS2092, Hytec) which provides event time stamps. When a fluorescence telescope in the adjacent TA building is triggered by a candidate UHECR shower, an external trigger is issued to the FAST DAQ with a typical rate of $\sim 3$ Hz. A trigger initiates the capture of a 4000 bin frame of data, corresponding to 80 $\mu$s.

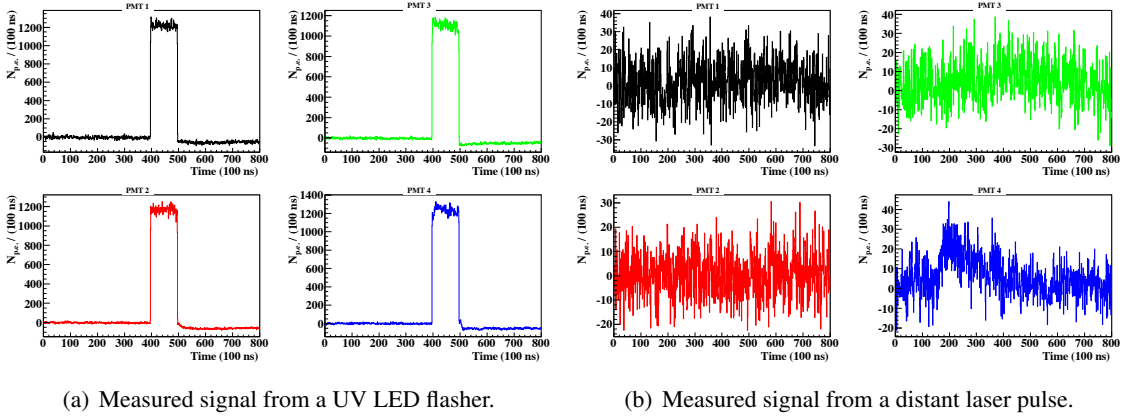

(a) Measured signal from a UV LED flasher.     (b) Measured signal from a distant laser pulse.

**Figure 3:** Example waveforms recorded with the full-scale FAST prototype. (a) Signal in each of the four PMTs following illumination of the telescope aperture with a UV LED flasher. (b) Signal from a vertical UV laser pulse at a distance of 21 km.

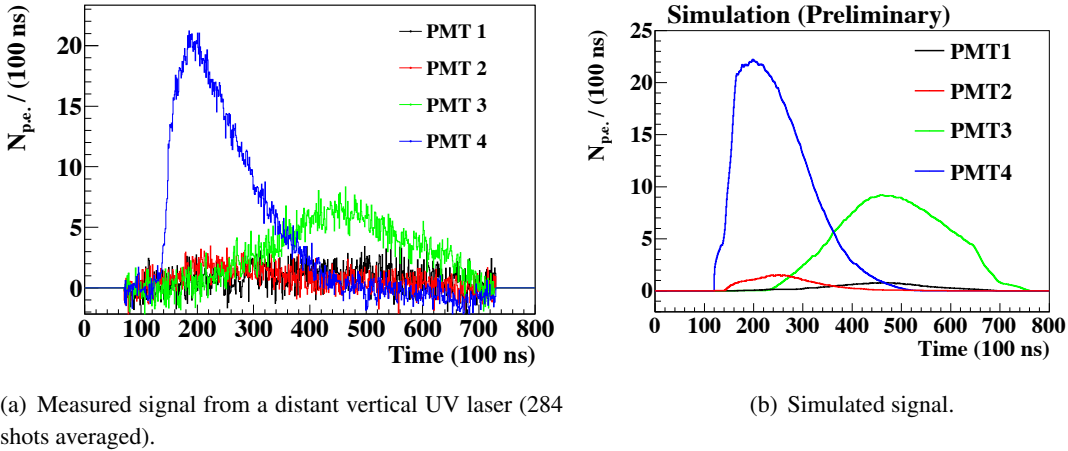

(a) Measured signal from a distant vertical UV laser (284 shots averaged).     (b) Simulated signal.

**Figure 4:** (a) Average waveform from 284 vertical UV laser pulses at a distance of 21 km and (b) simulated waveform from a raytracing simulation (without night sky background included).

Figure 3(a) shows example waveforms recorded with the FAST DAQ from a UV LED flasher illuminating the telescope aperture, confirming the timing synchronization between the signals, as well as the uniformity in the PMT gains. The measured signal was converted to a photo-electron count using a calibration factor measured in the laboratory. Figure 3(b) shows the measured signal from a 355 nm vertical UV laser at a distance of 21 km [10]. The laser signal, with a nominal pulse energy of 4.4 mJ, is approximately equivalent in intensity to a $\sim 10^{19.5}$ eV UHECR at a distance of 21 km.

Figure 4(a), the 284 waveforms of the vertical UV laser were averaged to improve the signal-to-noise ratio. As a result of the trigger algorithm used by the TA FD, the leading edge of the laser signal in each waveform has to be adjusted based on the GPS timing. Figure 4(b) shows the expected signal from a distant vertical laser evaluated from angular responses described as below, assuming typical atmospheric attenuation properties. The expected signal is in good agreement with observations. Small differences between the simulated and measured waveforms can be explained by the non-uniformity of the PMT surface, and the temperature dependence of the PMT gains.

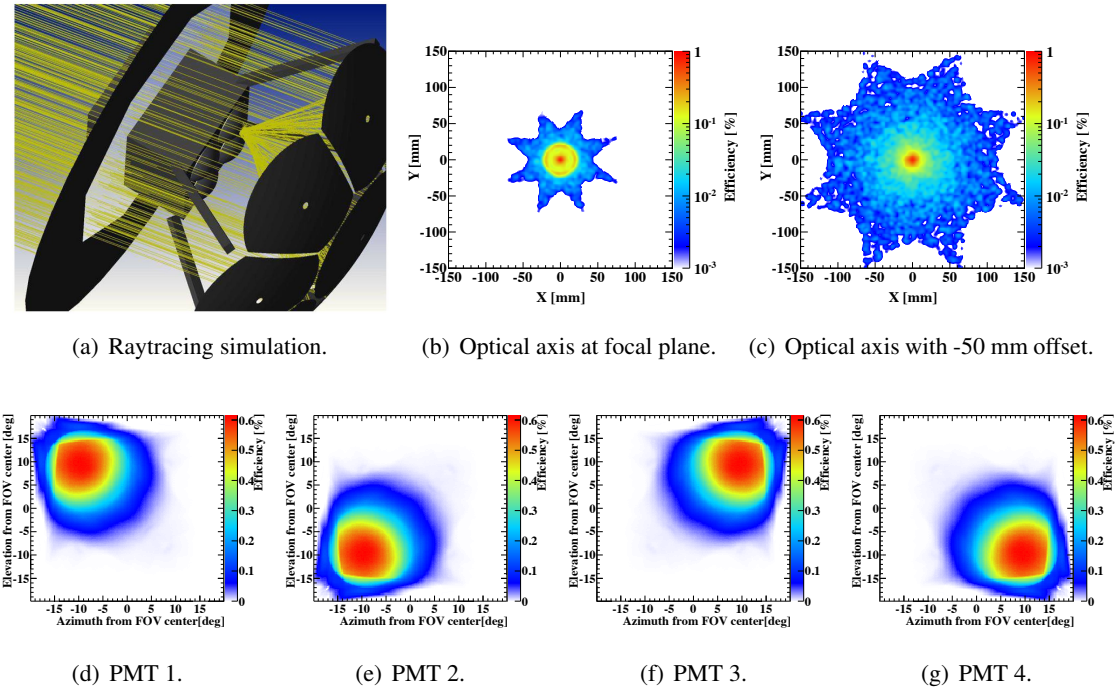

(a) Raytracing simulation.   (b) Optical axis at focal plane.   (c) Optical axis with -50 mm offset.

(d) PMT 1.   (e) PMT 2.   (f) PMT 3.   (g) PMT 4.

**Figure 5:** (a) Schematic view of a raytracing simulation of the full-scale FAST prototype optics. The spot size at the focal plane (b) and with a -50 mm offset (c) (for a beam parallel to the optical axis of the telescope) [8]. The angular efficiency characteristics of PMTs 1-4 are indicated in (d)-(g).

A raytracing simulation was performed to determine the optical characteristics of the prototype. A parallel beam of photons is injected at the telescope aperture, and individual photon paths through the telescope are calculated as shown in Figure 5(a). Figure 5(b) and (c) show the spot size at the focal plane, and with a -50 mm offset (which is of relevance as the PMTs installed in the camera have a spherical surface). Additional information about the optical characteristics of the prototype telescope is presented at this conference [8]. Considering these characteristics, angular responses for all four PMTs are evaluated as shown in Figure 5(d) to (g).

## 4. UHECR shower search

We continue steady operation of the full-scale FAST prototype, with the system being fully remotely operable. As of May 2017, the total operation time is reached $\sim 150$ hours. In this dataset, UHECR shower signals are searched for via coincidences in more than 2 PMTs, after excluding airplane events and measurements of the vertical laser. 18 significant shower signals have been found in time coincidence with TA FD reconstructed events. Figure 6(a)(b) are two of the UHECR events observed with the new FAST prototype. The reconstructed energies from the TA FD monocular analysis [11] are $10^{18.08}$ eV at an impact parameter $R_p = 2.4$ km and $10^{18.55}$ eV at $R_p = 3.0$ km respectively.

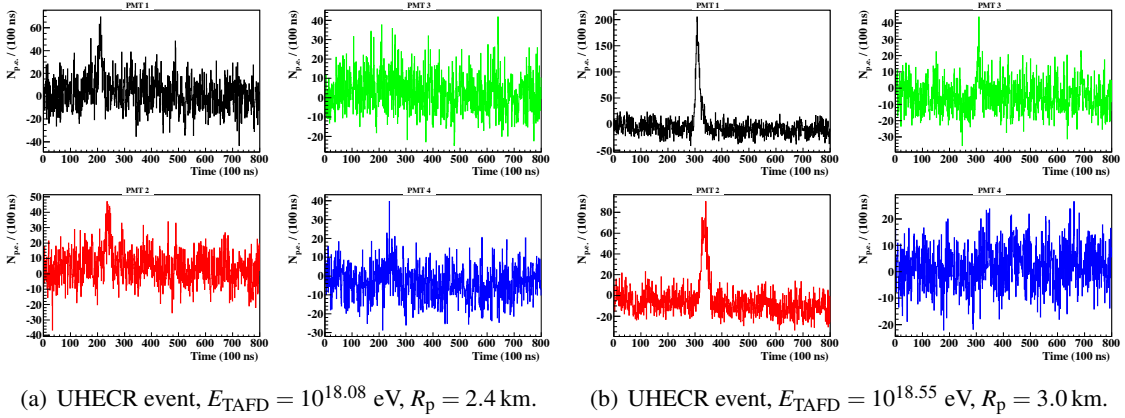

(a) UHECR event, $E_{\text{TAFD}} = 10^{18.08}$ eV, $R_p = 2.4$ km.  (b) UHECR event, $E_{\text{TAFD}} = 10^{18.55}$ eV, $R_p = 3.0$ km.

**Figure 6:** First UHECR signals measured with the full-scale FAST prototype.

## 5. Summary and Future Plans

We have presented a novel concept for a next-generation fluorescence detector, which features just a few pixels covering a large field-of-view. In October 2016, we installed the full-scale FAST prototype at the Telescope Array site in central Utah, USA, and began data taking in coincidence with the TA fluorescence detector. The prototype was used to measure the signal from a UV LED flasher, as well as a distant vertical UV laser beam. We were also successful in detecting 18 UHECR events in time-coincidence with the Telescope Array fluorescence detector. We will continue to operate the prototype and search for UHECR events.

## Acknowledgements

This work was supported by the Japan Society for the Promotion of Science through the Grant-in-Aid for Young Scientist (A) 15H05443, Grant-in-Aid for JSPS Research Fellow 16J04564 and JSPS Fellowships H25-339, H28-4564. This work was partially carried out by the joint research program of the Institute for Cosmic Ray Research (ICRR), University of Tokyo. This work was supported in part by NSF grant PHY-1412261 and by the Kavli Institute for Cosmological Physics at the University of Chicago through grant NSF PHY-1125897 and an endowment from the Kavli Foundation and its founder Fred Kavli. The Czech authors gratefully acknowledge the support of the Ministry of Education, Youth and Sports of the Czech Republic project No. LG15014, LE13012, LO1305, LM2015038, EU/MSMT CZ.02.1.010.00.016_0130001402.

The authors thank the Telescope Array Collaboration for providing logistic support and part of the instrumentation to perform this measurements. They also thank the Pierre Auger Collaboration for fruitful discussions.

# The Prototype Opto-mechanical System for the Fluorescence detector Array of Single-pixel Telescopes

**Dusan Mandat**[*,a], **Miroslav Palatka**[a], **Miroslav Pech**[a], **Petr Schovanek**[a], **Petr Travnicek**[a], **Libor Nozka**[b], **Pavel Horvath**[b], **Miroslav Hrabovsky**[b], **Justin Albury**[c], **Jose A. Bellido**[c], **John Farmer**[d], **Toshihiro Fujii**[e,†], **Aygul Galimova**[d], **Max Malacari**[d], **Ariel Matalon**[d], **John N. Matthews**[f], **Maria Merolle**[d], **Xiaochen Ni**[d], **Paolo Privitera**[d], **Stan B. Thomas**[f] **(FAST Collaboration)**[‡]

[a]*Institute of Physics of the Academy of Sciences of the Czech Republic, Prague, Czech Republic*
[b]*Palacky University, RCPTM, Olomouc, Czech Republic*
[c]*Department of Physics, University of Adelaide, Adelaide, S.A., Australia*
[d]*Kavli Institute for Cosmological Physics, University of Chicago, Chicago, IL, USA*
[e]*Institute for Cosmic Ray Research, University of Tokyo, Kashiwa, Chiba, Japan*
[f]*High Energy Astrophysics Institute and Department of Physics and Astronomy, University of Utah, Salt Lake City, UT, USA*
*E-mail:* mandat@fzu.cz

We present the opto-mechanical design of a new generation fluorescence telescope for the detection of ultrahigh-energy cosmic rays (UHECRs). The Fluorescence detector Array of Single-pixel Telescopes (FAST) is a proposed low-cost, large-area, next-generation experiment for the detection of UHECRs via the atmospheric fluorescence technique. The telescope is of a simplified Schmidt design, suitable for a camera consisting of only a few large pixels. The telescope has a 1 m$^2$ entrance aperture, and a field-of-view of $30° \times 30°$. We present the optical design of the prototype telescope as well as the mirror alignment and pointing calibration procedures. The prototype of the FAST telescope is installed at the Black Rock Mesa site of the Telescope Array Experiment.

[*]Corresponding author.
[†]Presenter.
[‡]https://www.fast-project.org

# 1. Fluorescence detector Array of Single-pixel Telescopes (FAST)

The Fluorescence detector Array of Single-pixel Telescopes (FAST) [1] is a design concept for a low-cost, ground-based fluorescence detector. A FAST telescope would consist of just four pixels covering a $30° \times 30°$ patch of the sky with a $\sim 1$ m$^2$ collecting area. Its low cost would facilitate deployment over a very large ground area, making it a viable candidate for a next-generation cosmic ray observatory. Such a design comes at the expense of low energy performance, as the signal-to-noise (S/N) ratio measured by a photomultiplier-tube (PMT) is proportional to $\sqrt{A/\Delta\Omega}$ [2], where $A$ is the light collecting area and $\Delta\Omega$ is the pixel solid angle, which is $\sim 15°$ in the FAST design (compared with, for example, $A \sim 3$ m$^2$ and $\Delta\Omega \sim 1.5°$ for the fluorescence telescopes of the Pierre Auger Observatory [3]). In addition, reconstruction of the geometry of an EAS with adequate resolution using data collected by a single FAST telescope is unlikely, as the coarse granularity of a $2 \times 2$ matrix of PMTs does not supply sufficient timing information to remove degeneracy in the determination of the shower axis. However, showers of sufficiently high energy would be observed by multiple FAST telescopes in an array, in which case timing information from the involved telescopes, along with the shape of the detected light pulse, could allow for reconstruction of the shower geometry with reasonable accuracy. An array of FAST telescopes would also be well suited as a complementary fluorescence detector to a sparse array of ground-based particle detectors, which could supply the shower geometry independently.

# 2. FAST prototype optical design

The prototype utilizes a large segmented mirror telescope of 1 m$^2$ collecting area to focus light onto a camera consisting of four 200 mm diameter PMTs. The prototype has been installed in a dedicated building alongside the fluorescence telescopes at the Black Rock Mesa site of the Telescope Array Experiment (TA) [4], where the design is being tested. Two additional FAST telescopes will be installed at the same location in September 2017, covering a total of $90°$ in azimuth. The deployment of additional FAST telescopes will allow for a three-fold increase in the prototype aperture, greatly increasing the number of showers observed in coincidence with the TA fluorescence telescopes. The primary design goal of an individual FAST telescope is to have an optical system with an effective collecting area of $\sim 1$ m$^2$ and a $\sim 30° \times 30°$ field-of-view which is capable of focusing atmospheric fluorescence light onto a matrix of several 200 mm PMTs. In addition, it is required to be low-cost, straightforward to maintain, and easy to transport and install. The low cost requirement makes it necessary to minimize the number of optical elements in the telescope. A Schmidt type optical design was adopted for the full-size FAST prototype. In a typical Schmidt telescope a corrector plate is placed at the entrance aperture (located at the mirror's radius of curvature, a distance of $2f$, where $f$ is the focal length) to facilitate the control of off-axis aberrations: coma and astigmatism. Field curvature and spherical aberration are still present, although the former can be eliminated by placing a suitably curved detector in the image plane. The size of the optical point spread function (PSF), which describes the spatial distribution of light on the focal surface, is a function of the spherical aberration of the system, and is typically circular in shape for both on- and off-axis beams.

The coarse granularity of the FAST camera, having only four PMTs each covering an angular field-of-view of $\sim 15°$, allows the requirements on the size and shape of the telescope's PSF to be relaxed. The FAST prototype telescope therefore takes the form of a lensless Schmidt camera, as residual coma and astigmatism present due to the lack of a corrector plate does not affect the functionality of the telescope. The telescope mirror is reduced in size, and the distance between the mirror and the focal surface shortened relative to a regular Schmidt telescope, with the entrance aperture located closer to the focal surface. The dimensions of the FAST prototype telescope are shown in Fig. 1. An octagonal aperture of height 1.24 m is located at a distance of 1 m from a 1.6 m diameter spherical mirror. The design fulfills the basic FAST prototype requirements, with an effective collecting area of 1 m$^2$ after accounting for the camera shadow, and a field-of-view of $30° \times 30°$. The size and the shape of the spot is of particular importance, and is shown in

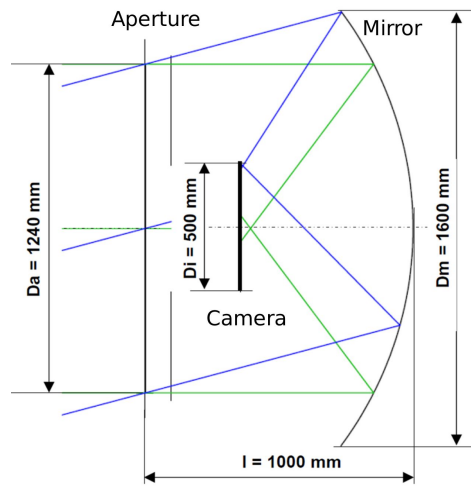

**Figure 1:** The dimensions of the FAST prototype telescope's optical system. $D_a$ is the face-to-face size of the octagonal telescope aperture, $D_i$ is the side length of the square camera box, $D_m$ is the diameter of the primary mirror, and $l$ is the mirror-aperture distance.

Fig. 2 (note that this figure depicts the geometrical spot shape along with the intensity of the spot relative to the maximum). The top (bottom) row shows the spot shape for an on-axis (off-axis) optical beam as a function of the distance from the focal plane. The 300 mm scale represents the maximum diameter of the spot size, the PMT diameter is 200 mm and 4 of them will be installed in the camera located in a custom-built box at the focus of the optical system. The characteristic "star" shape of the optical spot is a result of the octagonal shape of the entrance aperture. The spot shape becomes circular in nature for positive defocusing of the telescope (the image plane moved closer to the mirror), with a central hole corresponding to the shadow of the camera box. In order to minimize the effect of the dead space between PMTs, a 25 mm negative defocusing was utilized in the prototype design. This serves to eliminate a complete loss of signal for on-axis optical beams where light is focused in the central point between all four PMTs. These simulations were performed assuming a single compact primary mirror, while the constructed FAST prototype uses a segmented primary mirror, complicating the shape of the optical spot. Nevertheless, these simple simulations accurately predict its size.

The FAST telescope design consists of a central circular mirror and 8 side mirrors, or "petals".

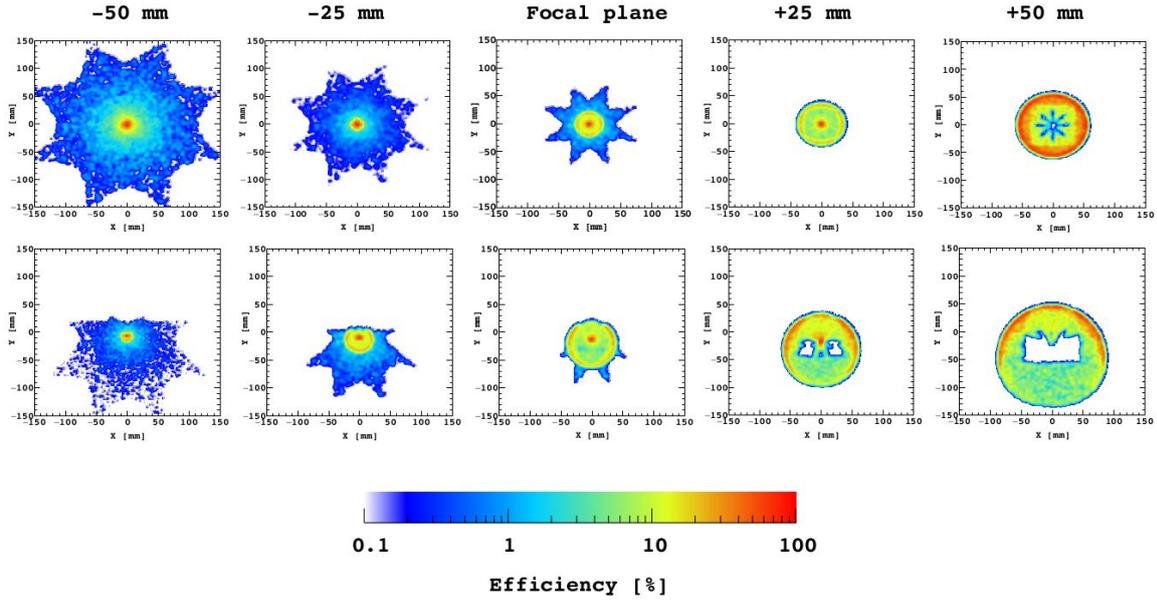

**Figure 2:** Geometrical spot diagrams from ray-tracing simulations of the FAST prototype optics. The spot diameters are shown for both on-axis (top) and off-axis (bottom) beams as a function of the distance from the focal plane (defocusing). Negative defocusing corresponds to moving the focal plane further from the mirror, and vice versa for positive defocusing. The color scale represents the relative intensity profile of the spots.

The diameter of the individual mirror segments was limited by the technology available in our laboratory. The mirrors are produced on site, in the Joint Laboratory of Optics of the Palacky University, and the Institute of Physics of the Academy of Sciences of the Czech Republic, from a custom-made substrate. The substrate is a borosilicate glass with good optical and mechanical quality. The reflective surface consists of vacuum coated Al and $SiO_2$ layers. The typical spectral reflectance, filter transmission and total optical efficiency between 260 nm and 420 nm is shown in Fig. 3. The reflectivity is relatively constant over this wavelength range, with a maximum of $\sim 90\%$ at 420 nm, and a minimum of $\sim 75\%$ at 260 nm.

A UV band-pass filter is installed at the aperture of the telescope to reduce the exposure to night-sky background light. We use a ZWB3 filter manufactured by Shijiazhuang Zeyuan Optics. Its spectral transmission is shown in Fig. 3.

The telescope's mechanical support structure was built from commercially available aluminum profiles. This allows for straightforward assembly/disassembly, and easy packing and transport due to their light weight, while also providing an extremely stable and rigid platform for the FAST optical system to be mounted on. The mechanics consists of a primary mirror stand mounted with a single degree of freedom to facilitate adjustment of the telescope's elevation (the elevation can be set to discrete values of $0°$, $15°$, $30°$ and $45°$ above the horizon). The square camera box (side length 500 mm), which holds four 200 mm PMTs, is mounted on a support structure connected to the perimeter of the mirror dish which also holds the octagonal filter aperture. The mirror stand contains 9 mirror mounts, each with 2 degrees of freedom to allow for mirror segment alignment.

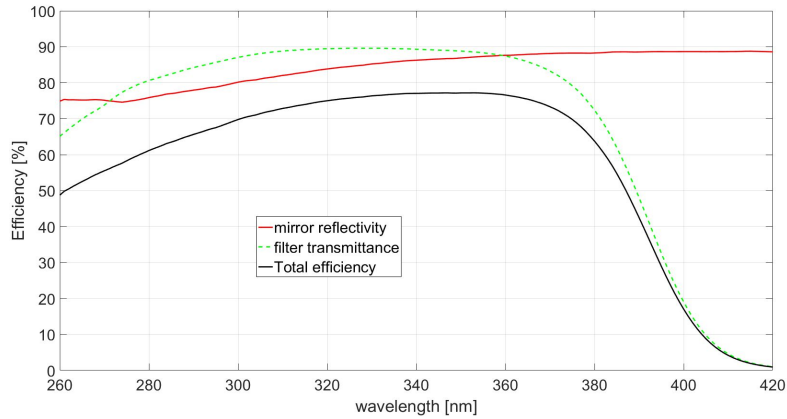

**Figure 3:** The typical spectral reflectance of the FAST mirror between 260 nm and 420 nm, along with the spectral transmission of the UV band-pass filter. The resultant total optical efficiency is shown in black.

The whole mechanical construction, shown in Fig. 4, is covered with a shroud to protect the optical system from the surrounding environment.

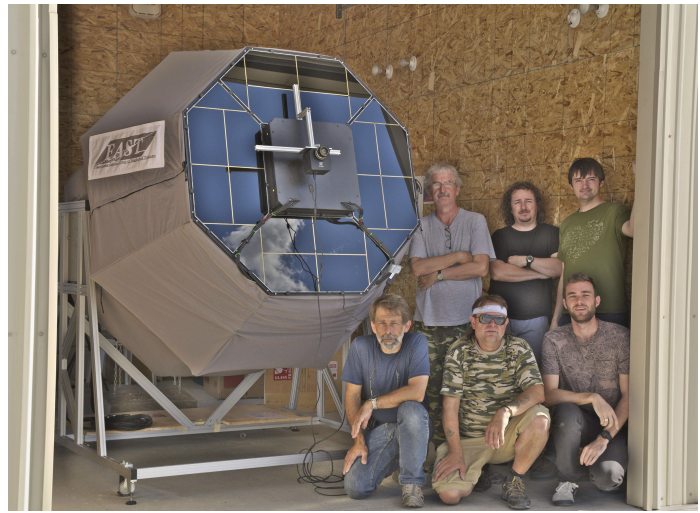

**Figure 4:** The complete mechanical structure. The mirror and the camera are protected from dust and aerosols by a shroud, which also acts as a shield for ambient light. The image was taken at the Black Rock Mesa site of TA following the installation.

## 3. Mirror alignment

The geometrical optical axis is defined by the line joining the center of the mirror dish to the center of the camera box. Once the geometrical axis of the FAST mirror dish is defined, all mirror segments must be aligned with respect to this axis. Each mirror is mounted on a support with two degrees of freedom for segment alignment, and a diode laser is mounted in the center of the mirror dish pointing along the optical axis. We use two light sources and two alignment techniques to accurately determine the final position of the mirrors. As the segmented mirror has a spherical shape, all mirror segments must point to the radius of curvature of the telescope. The light source is located along the optical axis of the telescope, at a distance of $2f$ from the central mirror. The point-like source is reflected back to a screen mounted at the same location, and if the segments are well aligned the image of the light spot overlaps the light source. The second spot-like source is placed close to the telescope on the optical axis, at a distance $< f$. The light is reflected by the mirrors creating a Bokeh image on a screen located on the optical axis at a distance $> 2f$. The Bokeh image creates a compact group of spots on the screen, allowing for fine-tuning of the mirror alignment. The alignment setup, along with images of the $2f$ and Bokeh alignment spots, are shown in Figure 5.

## 4. Pointing of the FAST telescope

The alignment of the telescope axis is of vital importance to the operation of the telescope. We use the laser (mentioned in section 3) to define the optical axis of the telescope. An astronomical camera with a fast f-number and an angular FOV of approximately 15° is located on the back cover of the FAST camera as close to the optical axis as possible. A screen is placed far from the telescope along the optical axis (the laser spot is located on the screen). The center of the image of the astronomical camera is overlapped with the laser light spot on the screen (the astronomical camera holder can be tilted to align its optical axis with that of the FAST telescope). The distance between the screen and the telescope defines the angular uncertainty in the alignment of the FAST telescope axis. The misalignment is negligible $< 0.01°$ for telescope-screen distances $> 200\,\mathrm{m}$. Once the alignment of the optical axis of the astronomical camera is complete, an image of the night sky can by obtained. We use the `astrometry.net` software package to calculate the right ascension and declination of the image center. The result can be transformed into the azimuth and elevation of the FAST telescope. The uncertainty in this method is very small, typically a few arcseconds.

## 5. Summary and Future Plans

We have presented a novel concept for a next-generation lensless Schmidt fluorescence telescope, which features just a few pixels covering a large field-of-view. The first full-scale FAST telescope prototype was installed in October 2016 at the Telescope Array site in central Utah, USA. The prototype was tested using observations of a UV LED flasher, as well as a distant vertical UV laser beam. It has also been used to detect UHECR events in time-coincidence with the Telescope Array fluorescence detector. A second prototype will be installed in October 2017 to increase the angular coverage and allow for the detection of more UHECR events.

## Acknowledgements

This work was supported by the Japan Society for the Promotion of Science through the Grant-in-Aid for Young Scientist (A) 15H05443, Grant-in-Aid for JSPS Research Fellow 16J04564 and JSPS Fellowships H25-339, H28-4564. This work was partially carried out by the joint research

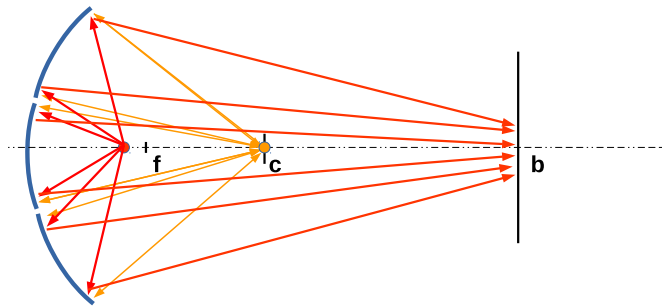

(a) The mirror alignment setup. Two light sources are used. Red for Bokeh and yellow for $2f$ alignment.

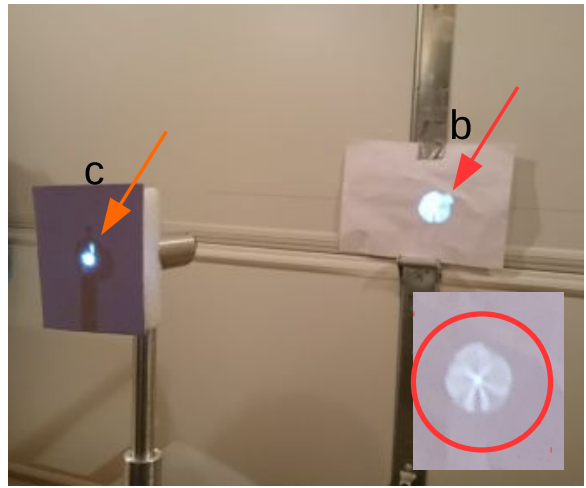

(b) The Bokeh screen (right) showing the alignment procedure. The (left) screen with a built-in light source is used for $2f$ alignment. $f$ is the focal plane position, $c$ radius of curvature of the mirror segments and $b$ is the location of Bokeh screen.

**Figure 5:** Mirror alignment, the misaligned mirror segment is visible on both screens - top right close to the spot center (see the arrow). The detail in the Bokeh of the aligned mirror is shown in the red circle (right bottom).

program of the Institute for Cosmic Ray Research (ICRR), University of Tokyo. This work was supported in part by NSF grant PHY-1412261 and by the Kavli Institute for Cosmological Physics at the University of Chicago through grant NSF PHY-1125897 and an endowment from the Kavli Foundation and its founder Fred Kavli. The Czech authors gratefully acknowledge the support of the Ministry of Education, Youth and Sports of the Czech Republic project No. LG15014, LE13012, LO1305, LM2015038, LTAUSA17078, EU/MSMT CZ.02.1.010.00.016_0130001402.